\documentclass{eas}
\usepackage{graphics}
\usepackage{graphicx}
\begin{document}

\title{Formation Efficiencies of Old Globular Clusters - form dwarf to giant galaxies}
\runningtitle{Formation Efficiencies of Old Globular Clusters}
\author{Iskren Georgiev}\address{Argelander Institut f\"{u}r Astronomie de Universit\"{a}t Bonn, Auf dem H\"{u}gel 71, 53225 Bonn, Germany, e-mail: iskren@astro.uni-bonn.de}
\author{Thomas Puzia}\address{Departamento de Astronomia y Astrofisica, Avenida Vicuna Mackenna 4860, 7820436 Macul, Santiago de Chile}
\author{Paul Goudfrooij}\address{Space Telescope Science Institute, 3700 San Martin Drive, Baltimore, MD 21218, USA}
\author{Michael Hilker}\address{European Southern Observatory, Karl-Schwarzschild-Str.\,2, 85748 Garching bei M\"{u}nchen, Germany}
\begin{abstract}
For the full galaxy mass range, we find that previously observed trends of  globular cluster (GC) system scaling parameters (number, 
luminosity or mass of all GCs in a galaxy normalized to the host galaxy luminosity or mass, e.g. $S_L$) as a function of galaxy mass, 
holds irrespective of galaxy type or environment. The $S_L$ value of early-type galaxies is, on average, twice that of late-types. We derive 
theoretical predictions which describe remarkably well the observed GC system scaling parameter distributions given an assumed GC 
formation efficiency ($\eta$), i.e. the ratio of total mass in GCs to galaxy halo mass. It has a mean value of $\eta\simeq 5.5 
\times10^{-5}$, and increasing scatter toward low galaxy mass. The excess $\eta$-values of some massive galaxies compared to 
expectations from the mean model prediction, may be attributed to an efficient GC formation, inefficient production of field 
stars, accretion of low-mass high-$\eta$ galaxies or likely a mixture of all these effects.
\end{abstract}
\maketitle
\section{Introduction}
The number, luminosity or mass of the entire globular cluster (GC) system normalized to the host galaxy luminosity or mass defines the 
fundamental quantities specific frequency ($S_N$), luminosity ($S_L$), specific mass ($S_M$) and specific number ($\hat{T}$) of GCs 
(cf. Eqns \ref{eqn:sn}). These GC scaling relations indicate how efficiently galaxies form GCs per unit of their luminosity or mass. Those 
has been observed to vary significantly, being high for dwarf and giant galaxies (\cite{Harris91, MillerLotz07, Peng08}), and with a 
minimum at a galaxy luminosity of $M_V\simeq-20.5$\,mag ($L_V\simeq10^{10}L_\odot$). That is, the two extreme galaxy mass 
regimes, dwarfs and giants, seemingly form old GCs in similar proportions. This scaling relations provide important observational 
constraints and test for models of star, GC and galaxy formation (\cite{McLaughlin99, Forbes05, Pipino07, Peng08, MuratovGnedin10}). 

\section{Analysis and Results}\label{Sect:Analysis}
In Georgiev et al.\,(2010), we investigate this trend with observations of GC populations of a large sample late-type dwarf galaxies with 
HST/ACS (\cite{Georgiev08, Georgiev09}). In order to sample the entire range in galaxy mass, environment, and morphology we augment 
our sample with data with data from the literature. We find that $\it i)$ relations between the GCS scaling parameters and galaxy 
luminosity holds irrespective of galaxy morphological type and $\it ii)$ on average, early-type galaxies have $\sim2\times$ higher 
$S_L$-values than late-types at the same luminosity. To investigate the observed trends, which have not yet been conclusively explained, 
we derive theoretical predictions of GC system scaling parameters as a function of the total host galaxy mass based on the models of 
Dekel\,$\&$\,Brinboim (2006) in which star-formation processes (i.e. thermal properties of the gas) are regulated by stellar/supernova 
feedback below a stellar mass of $3\times10^{10} M_\odot$, and by virial shocks ('hot stream') above it, causing a suppression of star 
formation.

\begin{figure}[h]
\resizebox{1.\columnwidth}{!}{
\includegraphics[bb= 14 30 720 284]{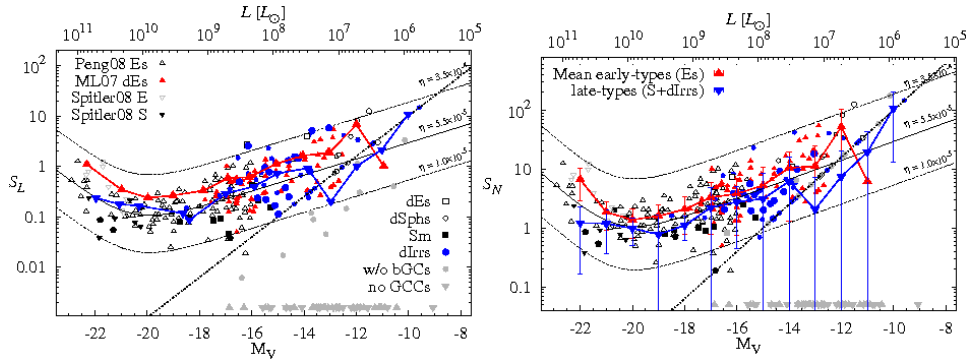}}
\caption{{\bf Left:} GC specific specific luminosities ($S_L$) and {\bf right} specific frequencies ($S_N$) as a function of galaxy 
luminosity. Solid curves are model predictions (cf. Eqs \ref{eqn:snn}) reflecting the different M/L below and above ${\cal M }\sim 3\times 
10^{10}M_\odot$. Solid curve is the best-fit $\eta_L=5.5\times10^{-5}$ to the $S_L$-value. Solid symbols connected with line show the co-added running average $S_N-$values per magnitude bin for early- and late-type galaxies. Dash-dotted line indicates the $S_N-$value if a galaxy has one GC (Eq.\ref{eqn:sn}). Grey triangles at the bottom of the plot represent galaxies for which no GC candidates were detected.
}\label{fig:2}
\end{figure}

Specific frequency and luminosity
\begin{eqnarray}
S_N=N_{\rm GC}\times10^{0.4(M_V+15)}; \ S_{L,V} = 10^2\times L_{\rm GCS}/L_V\label{eqn:sn}
\end{eqnarray}
can be related to the galaxy halo mass by defining the GC formation efficiency: $\eta = {\cal M}_{\rm GCS} / {\cal M}_h$.
Dekel \& Birnboim (2006) model predicts $L_V = {\cal M}_h^{5/3}$ below and $L_V = {\cal M}_h^{1/2}$ above galaxy stellar mass $\sim 
3\times 10^{10}M_\odot$.
\begin{eqnarray}
S_N = 10^{6+0.4M_{V,\odot}}\frac{\eta}{m_{\rm TO}}\left (\kappa_1^{0.6}L_V^{-0.4} +\kappa_2^{2}L_V^{1.2}\right)\label{eqn:snn};\ 
S_L = 10^2\frac{\eta}{\gamma_V}\left(\kappa_1^{0.6}L_V^{-0.4} + \kappa_2^{2}L_V\right),
\end{eqnarray}
where $\kappa_1=10^9\ {\rm and }\ \kappa_2=10^{-4.42}$ are observationally derived normalization constants, $m_{\rm 
TO}\simeq2\times10^5M_\odot$ and $\gamma_V=2$ are the typical GC mass and mass-to-light ratio. The functional relations (Eqs 
\ref{eqn:snn}) describe remarkably well the observed distributions (Fig.\,\ref{fig:2}). This supports that GCs form in proportion to the 
strength of the host potential (galaxy mass) as well as the effect on the GC formation efficiency by the physical mechanisms in the Dekel 
\& Birnboim model of galaxy evolution (thermal properties, shock stability and cooling physics of gas due to feedback from SNe, UV on 
dust, photoionization, AGN and dynamical friction). Therefore, this model is a good representation of the GC formation efficiency 
($\eta$). A better model should, in addition, take into account variations in the individual galaxy star formation history, merging history, 
conditions for cluster formation and destruction and the stochastic nature of star formation at low galaxy mass. Observations of GCSs of 
late-type, spiral galaxies in the high galaxy mass regime are necessary to probe whether their GC formation efficiencies are statistically 
different than that of early-type galaxies at same galaxy mass.


\end{document}